\documentclass[twocolumn,twoside,showpacs,preprintnumbers,amsmath,amssymb]{revtex4}

\usepackage{graphicx}
\usepackage{dcolumn}
\usepackage{bm}

\newcommand {\bc} {\begin{center}}
\newcommand {\ec} {\end{center}}
\newcommand {\be} {\begin{equation}}
\newcommand {\ee} {\end{equation}}

\newcommand {\dx} {\displaystyle}

\begin{document}

\title{Quantum Diffusion}

\author{Dimitar I. Pushkarov}
 \affiliation{Institute of Solid State Physics
 Bulgarian Academy of Sciences \\
 1784 Sofia, Bulgaria}

\date{\today}

\begin{abstract}
Basic ideas and results which characterize quantum diffusion of
defects in quantum crystals like solid helium as a new phenomenon
are presented.  Quantum effects in such media lead to a
delocalization of point defects (vacancies, impurities etc.) and
they turn into quasiparticles of a new type --- defectons, which
are characterized not by their position in the crystal lattice but
by their quasimomentum and dispersion law. Defecton-defecton and
defecton-phonon scattering are considered and an interpolation
formula for the diffusion coefficient valid in all interesting
temperature and concentration regions is presented.  A comparison
with the experimental data is made. Some alternative points of
view are discussed in detail and the inconsistency of the
Kisvarsanyi-Sullivan theory is shown.
\end{abstract}

\pacs{Valid PACS appear here}
\maketitle

\section{\label{sec:Intro} Introduction}

The basic characteristic of a quantum crystal is the presence of a
large amplitude of atomic zero-point vibrations. This means that
the overlap of the wave functions of the nearest neighbours is not
negligible and leads to considerable tunnelling transitions to the
nearest lattice sites. If a point defect (vacancy, impurity,
interstitial) appears in such a system, the crystal state becomes
degenerate with respect to the defect position due to the identity
of lattice sites and the defect delocalizes and turns into a
quasiparticle - \textit{defecton}. If the wave function of the
crystal with a localized defect in the lattice site $\mathbf{r}$
is $\psi_\mathbf{r}(x)$, then the right wave function of the
crystal, according to the Bloch theorem, is \cite{JETP70}
\begin{equation}\label{eq:Bloch}
\dx \Psi_\textbf{k} = \frac{1}{\sqrt
N}\sum_\textbf{r}\psi_\textbf{r}(x) \exp\{i(\textbf{kr}-\epsilon
(\textbf{k})t/\hbar)\}
\end{equation}
\noindent where $x$ is a $3N$ component vector which consists of
all atomic coordinates, and \textbf{k} is the defecton
quasiwavevector. The spectrum is given by the defecton dispersion
law $\epsilon(\mathbf{k})$. In a tight binding approximation it
has the form
\begin{equation}\label{eq:dispersion}
\epsilon(\textbf{k}) = \Phi + A\omega(\textbf{k})
\end{equation}
\noindent where $\Phi $ is the activation energy of a localized
defect, $A$ is the tunnelling probability amplitude (exchange
integral), and $\omega(\textbf{k})$ is a dimensionless function
which depends on the symmetry of the crystal lattice. For a simple
cubic lattice the dispersion law is of the form
\begin{equation}\label{eq:cubic}
\epsilon(\textbf{k}) = \Phi  + A(\cos k_xa+\cos k_ya+\cos k_za)
\end{equation}
\noindent and for body-centered cubic lattice (b.c.c.)
\begin{equation}\label{eq:bcc}
\epsilon(\textbf{k}) = \Phi
+4A\cos{\frac{k_xa}{2}}\cos{\frac{k_ya}{2}} \cos{\frac{k_za}{2}}
\end{equation}
\noindent where $a$ is the lattice constant. For defects in solid
helium, the quantity $\Phi $ is of the order of several degrees.
In case of h.c.p. lattice, there are two atoms per unit cell and
the dispersion law consists of two branches called conditionally
\emph{acoustic} and \emph{optical} respectively
\cite{FNT75kinetics,Singapore,Nauka}:
\begin{eqnarray}\label{eq:hcp}
\epsilon(\textbf{k})_{\frac{ac}{opt}} = \epsilon_0 + A\left\{ 7
-4\cos{\frac{\mathbf{k_1.a_1}}{2}}\cos{\frac{\mathbf{k_2.a_2}}{2}}
\cos\frac{\mathbf{k_3.a_3}}{2} \right. \nonumber\\
\left. \mp \cos\frac{\mathbf{k.c}}{2}\left[ 1 +
8\cos{\frac{\mathbf{k_1.a_1}}{2}}\cos{\frac{\mathbf{k_2.a_2}}{2}}
\cos{\frac{\mathbf{k_3.a_3}}{2}}\right]^{1/2} \right\}
\end{eqnarray}
\noindent where
\begin{equation}
\mathbf{a_1} = a\left[1,0,0\right],  \mathbf{a}_{2,3} =
a\left[-\frac {1}{ 2},\pm \frac{\sqrt3}{ 2}, 0\right], \mathbf{c}
= a\left[0, 0, \frac{\sqrt8}{3}\right].
\end{equation}

An important characteristic of a defecton is its energy band width
$\Delta$. For simple lattices $\Delta = zA$, where $z$ is the
number of nearest neighbours in the lattice. For the h.c.p.
lattice, the dispersion law (\ref{eq:hcp}) is written in a form
which fixes the bottom of the acoustic band at $k=0$, so that
$\epsilon(0) = \epsilon_0 $ is the lowest band level. The band
width of the acoustic branch is $\Delta_{ac}= 7A$ while that of
the optical one it is $\Delta_{opt}= 5A$. The bottom of the
optical branch lies higher by $3A$. A typical value of $\Delta$
for $^3$He impurities in $^4$He is $\Delta\sim10^{-4}K$, while for
vacancions $\Delta$ is of the order of several degrees. The
defecton velocity is
 $ \dx
\textbf{v} = \frac{1}{\hbar} \frac{\partial
\epsilon}{\partial\textbf{k}}
 $
and may be estimated by the relation $v\sim a\Delta / \hbar$. The
effective mass for simple lattices is $m^* = \hbar^{2} /Aa^{2}$
while for the h.c.p. dispersion law (\ref{eq:hcp}) one has
$m^*_{hcp} = \hbar^{2} /2Aa^{2}$. Another important quantity is
the delocalization time $ \tau_ 0$ which may be estimated by the
relation \cite{Singapore,DelocPSS75}
\begin{equation}\label{eq:tau_0}
\tau_{0}\sim A/ \hbar .
\end{equation}
Note, that in this relation the characteristic energy is the
exchange integral $A$, not the band width $\Delta$.

Before going on with defecton diffusion let us note that there are
crystals in which zero-point vibrations are large only for certain
degrees of freedom and hence the quantum properties manifest
themselves with respect to them. The class of quantum crystals
includes therefore heavy metals (palladium, niobium etc.) in which
the behaviour of light hydrogen impurities (protons) is quantum,
solid hydrogen with impurities of hydrogen atoms, metallic
hydrogen, etc.

From a historical point of view, however, physics of quantum
crystals is above all bound up with solid helium. Helium has been
chosen for a basic system when examining the ideas and results of
the theory owing to a number of circumstances.  Solid helium is
the best investigated quantum crystal.  Helium crystals are
extremely pure -- all other elements solidify at higher
temperatures, so the only impurities are isotopes and can be well
controlled in wide concentration ranges. The two helium isotopes
obey two quantum statistics -- those of Bose-Einstein and
Fermi-Dirac. In addition, each isotope crystallizes in three
different modifications. In such a way \emph{solid helium offers
six quantum crystals, composed of two isotopes only}. Phase
diagrams for helium isotopes can be found, e.g. in
\cite{Esselson-book82,QuantumLiquidsBook,Silvera}.

Any correct theoretical model has to include not only a
description of the structure and building particles, but also the
typical kinds of motion in it, or, in other words, \emph{the
elementary excitations as elementary carriers of the motions}. In
this sense crystals of solid $^4$He with small con\-centrations of
isotopic impurities and vacancies may serve as examples of really
existing model systems, in which the only elementary excitations
are phonons and defectons. In addition, the low value of the
melting temperature compared to the Debye temperature leads to a
simple linear dispersion law for phonons in the whole range of
existence of the solid phase. It is worth noting that the new
(tunnel) kind of motion presented by the defecton quasiparticles
is the reason to call the state of quantum crystals the \emph{a
new state of the matter}.

For $^3$He atoms which have spin equal to $1/2$ and relatively
large magnetic moment an exchange of atoms is equivalent to a spin
exchange (the interatomic potential practically does not depend on
spin) and therefore may be observed experimentally, e.g. by means
of nuclear magnetic resonance (NMR) \cite{AbragamGoldman}.

The presence of two types of crystals composed of Fermi- and Bose-
atoms puts forward the problem of defecton statistics.  It can be
shown that in pure crystals with small defect concentration
vacancions obey the statistics of host atoms. Impurities of $^3$He
in crystals of $^4$He as well as impurities of $^4$He in crystals
of $^3$He turn into Fermi-quasiparticles. In case of mixed
crystals a vacancy has to exchange places with different atoms
($^3\mathrm{He}$ or $^4$He) that obey different statistics,
therefore, the vacancion statistics can differ from the known
statistics of Bose-Einstein or Fermi-Dirac. More information on
defecton statistics can be found elsewhere
\cite{Singapore,DefectonStatPSS86}. The degeneracy temperature
$T_0$ is extremely small:
\begin{equation}\label{eq:T_0}
T_0\sim\Delta x^{2/3}
\end{equation}
\noindent where $x$ is the fractional concentration of defects.
For vacancions with a typical concentration $x\sim10^{-3}$ the
degeneracy temperature is $T_0\sim10^{-2} - 10^{-3} \mathrm{K}$.
That is why in many cases defectons may be considered as a
Boltzmann gas. It is important to note, however, that the role of
statistics increases with decreasing concentration.  Indeed, the
interaction energy $E_{int}$ between point defects in solids is
inversely proportional to the third power of the spacing $r$ and
hence is proportional to the concentration $x$:
\begin{equation}\label{eq:E_int}
 E_{int} = \frac{V_0 a^3}{r^3} = V_0 x
\end{equation}
\noindent where $V_0$ is the elastic interaction constant which
typical value is of the order of $10^{-2}K$. Taking into account
(\ref{eq:T_0}), one obtains
\begin{equation}\label{eq:T_0/E_int}
\frac{T_0}{E_{int}} \sim \frac{\Delta}{V_0}x^{-1/3}.
\end{equation}
Therefore, at low enough concentrations the interaction energy $E$
turns out to be negligible compared to the ''quantum  statistics
energy``  $T_0$. The critical concentration for a gas of vacancies
is $\sim10^{-1}$ and is much greater than the usual vacancy
concentration in the experiments.

\section{ Historical Review}

A phenomenological theory of defects in quantum crystals was
proposed first by A.F.Andreev and I.M.Lifshitz \cite{AL69} in
1969. The qualitative estimations concerning diffusion in that
work differ from the quantitative ones more than $7$ orders of
magnitude and are only of historical interest now. Simultaneously,
the author of this work reported at the Second Soviet Conference
on Theory of Solid State in Moscow his microscopic theory of
vacancions and impuritons (published as a paper \cite{JETP70} in
1970) where the defecton parameters were estimated and the
temperature dependence of the defecton mean free path was
obtained.

At the same time Guyer and Zane \cite{GuyerZane70} published their
perturbation theory of the spectrum of impuritons (called
\emph{mass-fluctuation waves}) and showed the possibility of
experimental observation by the method of nuclear magnetic
resonance.

It is worth noting that there was a significant interest in
quantum crystals at those years but there were no experimental
evidences on quantum diffusion. Moreover, some results on the
diffusion of point defects \cite{GriJETP73,GriJETP74} were
considered as arguments in favour of the classical activation
mechanism. There was not any clear conception of the correlation
between the quantum and classical weights in the phenomena
considered. This prevented often even the correct formulation of
the problems. Instructive in this aspect is the work of C.P.Flynn
and A.M.Stoneham \cite{FlynnSton70} on the diffusion of light
interstitials. The authors have practically neglected the coherent
quantum diffusion stressing the role of phonons and deformation
around the defect. This was a result of their conception that the
incoherent (by means of phonons) transition corresponded to a
macroscopic number of channels, while the coherent tunnelling
corresponded to one. In this way, an implausible estimation has
been made for the temperature region in which coherent diffusion
could be observed, given by the inequality $T\leq10^{-17}K$.

The assumption that defects turn into quasiparticles means that
(at some conditions considered later on) the gas kinetic model is
applicable and hence the diffusion coefficient has to be inversely
proportional to the concentration, i.e.
\begin{equation}\label{eq:D}
D \sim Aa^2 / \hbar x \sigma
\end{equation}
\noindent where $\sigma$ is the defecton-defecton cross-section
(in units $a^2$). The temperature dependence predicted
\cite{PhD,JETPLett74} was of the form
\begin{equation}\label{eq:D_T}
D_T \sim as \left(\frac{\Delta}{\theta_p} \right)^2 \left(
\frac{\theta_p}{T} \right)^9, \quad \theta_p =\frac{\hbar s}{2a} .
\end{equation}
  The ninth power of temperature is easy to understand (for
details see Refs.~\onlinecite{JETP70,DiffusionJETP75}). The motion
of a defect is determined by the number of phonons ($\dx N_{ph}
\sim T^3$ at low temperatures) and, by the scattering of long wave
phonons on defects (Rayleigh scattering) with a cross-section
$\sigma\sim k^4 \sim T^{4}$; finally, the motion of a ''heavy and
slow`` defecton in a gas of ''light`` phonons depends on the
second power of the phonon wave vector ($\dx \sim k^2\sim T^2$).
Hence, $D\sim 1/(N_{ph}\sigma T^2) \sim T^{-9}$. It is worth
noting that the role of the characteristic temperature is played
by $\theta_p$, not by the Debye temperature $\theta_D$. These two
temperatures differ significantly: $\dx\theta_p = \frac{\hbar
s}{2a} \approx \frac{\theta_D}{8}$ and lead to a numerical
difference of the order of $\ 8^7 \sim 10^6$ times.

It could be expected that such a rapid increase of the diffusion
with decreasing temperature can easily be observed experimentally.
However, the experiments on both vacancy \cite{GriJETP73} and
impurity \cite{Miyoshi70} diffusion carried out in $^4$He with
$0.75\div 2\%$ impurity concentration indicated typical classical
exponential decreasing down to the phase separation temperature.

The first evidence of a quasiparticle behaviour was reported by
R.G.Richards, J.Pope and A.Widom \cite{RichardsPopeWidom72}. They
confirmed the concentration dependence $\sim 1/x$.  However for
the tunnelling frequency $A$ they obtained the value
$A\sim10^{-7}K$, which is three orders of magnitude smaller than
the exchange integral $J_{33}$ in pure $^3$He.  The interpretation
of such a value ran into difficulties because there were no
physical reasons for such dramatic difference in the exchange
integrals. It became clear that additional information is
necessary and that it could be obtained from the temperature
dependence (\ref{eq:D_T}).  An increasing diffusion with
decreasing temperature was first observed by the Kharkov group
\cite{GriPismaJETP73,GriJLTP73} in 1973.  Surprisingly, the
treatment of the experimental data using the theory of Andreev and
Lifshitz \cite{AL69} gave for the exchange integral $A$ a value of
the same order of magnitude as that obtained in
\cite{RichardsPopeWidom72}.

The plain model of free quasiparticles became questionable and in
1974 Landesman and Winter \cite{LandesmanWinterLT13} and later on
Huang et al. \cite{Huang-Guyer75} proposed an alternative model
leading to another concentration dependence:
\begin{equation}\label{eq:D_Landesman}
D\sim\frac{Ja^{2}}{\hbar x^{4/3}}, \qquad J = \frac{J'^2}{K}
\end{equation}
\noindent where $J'$ is the exchange integral in the pure crystal,
and $K$ is a phenomenological parameter describing the repulsion
of two atoms in one lattice site.

The authors argued that this dependence is valid for
concentrations $x >10^{-6}$, i.e.  in the whole experimentally
accessible region. Unfortunately, the experimental precision was
not enough to distinguish the small difference between the
concentration dependencies in  (\ref{eq:D}) and
(\ref{eq:D_Landesman}), especially taking into account that both
$A$ and $J$ were unknown.

In order to solve the paradox the author treated the experimental
data of the Kharkov group using his theory \cite{JETPLett74}. As a
result, the exchange integral $A$ was found to be only slightly
smaller than $J_{33}$, the band width was estimated to be
$\Delta\sim10^{-4}K$ and the prediction of the giant
defecton-defecton cross-sections in \cite{JETP70} was confirmed.
The existence of the new quasiparticles (defectons) became an
immutable fact and the discovery of a new phenomenon called
\emph{quantum diffusion} was acknowledged.

\section{Defecton-defecton scattering}

The only scattering processes which determine the defecton
diffusion are collisions with defectons and phonons. If the
temperature is small enough (the exact conditions will be derived
below) the number of phonons is negligible, and the defecton free
path is controlled by defecton-defecton interaction.

Defecton-defecton scattering possesses a series of peculiarities
relevant to the energy band width
\cite{JETP70,JETPLett74,DiffusionJETP75,Singapore}. First of all,
the quasiparticle energy cannot vary during the motion more than
the energy band width $\Delta$. This means that the quasiparticle
is not able to overcome barriers higher than $\Delta$. In case of
impuritons in helium, a typical value of the deformation energy
caused by a vacancy or impurity is of the order of
$10^{-2}~\mathrm{K}$ and is much larger than $\Delta_{imp} \sim
10^{-4}$~K. As a result, the quasiparticle cannot reach the core
of the vacancy (or interstitial) and exchange place with it (the
later case of vacancy-stimulated diffusion has been discussed in
Sec.~\ref{sec:Locke}). This is important because the exact form of
the deformation in the core is complicated and not known, while
the behaviour of the deformation field at large distances is
common for wide classes of solid media. The later allows to study
the defecton scattering in a quite general form.

Point defects in crystals interact by the law
\begin{equation}\label{eq:fi_anisotrop}
\varphi(\textbf{r}) = V_0 \Gamma(\textbf{n})\left( \frac {a}{r}
\right)^3 ,
\end{equation}
\noindent where $\Gamma(\textbf{n})$ is a function of the
direction $\textbf{n}$, which vanishes after angle averaging and
is of the order of unity.  In an isotropic media
\begin{equation}\label{eq:fi_isoprop}
\varphi(r) = V_0 \left(\frac {a}{r} \right)^6 .
\end{equation}
Let us consider first the features of the defecton scattering in a
potential of a more general form:
\begin{equation}\label{eq:fi_general}
\varphi(r) = V_0 \left( \frac ar \right)^n .
\end{equation}
The small value of the impuriton energy band width leads to a
peculiar confinement in areas with strong field gradients. This
effect takes place both in attractive and in repulsive potentials.
Its physical reason is that the transition of the quasiparticle
into a neighbouring lattice site cannot take more energy than the
band width.  Therefore, the condition for such a confinement may
be written in the form
\begin{equation}\label{eq:local-condition}
\mid\nabla\varphi\mid a >\Delta .
\end{equation}
Substituting here $\varphi$ from  (\ref{eq:fi_general}) and
replacing the mean distance $r$ between defects by their
concentration $x\approx(a/r)^3$  one obtains that all defects has
to be localized \cite{AL69}  at concentrations $x>x_c$, where
\begin{equation}\label{eq:x_c}
x_c  = \left( \frac{\Delta}{nV_0} \right)^{\frac{3}{1+n}} .
\end{equation}
In the case of isotopic impurities in helium $(n=3)$ the quantity
$x_c$ is
\begin{equation}\label{eq:x_c-value}
x_c = \left( \frac{\Delta}{3V_0} \right)^{\frac{3}{4}} \approx
1\div 2\% .
\end{equation}
Due to the small value of the defecton band width large regions
around the impurities are unscalable for quasiparticles
\cite{JETP70,DiffusionJETP75}. Their linear size $R_0$  may be
estimated from the equation

\begin{equation}\label{eq:R_0}
\varphi(R_0) =\Delta , \quad R_0 =
a\left(\frac{V_0}{\Delta}\right)^{1/n} .
\end{equation}

Hence, the effective cross-section $\sigma$ for impuritons in the
potential (\ref{eq:fi_general}) is
\begin{equation}\label{eq:sigma}
\sigma \approx \pi a^2\left(\frac{V_0}{\Delta}\right)^{2/n} .
\end{equation}
These estimations may be improved taking into account, that the
experiments are performed at temperature $T~\gg~\Delta$ and,
therefore, the conditions for the quasi-classical approximation
are satisfied:
\begin{equation}\label{eq:quasi-class}
(ka)^{n-2} \gg\frac{A}{V_0} \approx\frac{\Delta}{zV_0} .
\end{equation}
In this case the cross-section is given by the expression
\cite{JETP70}:
\begin{equation}\label{eq:sigma-quasi}
\sigma \approx 2\pi
a^2\left(\frac{V_0}{Aka}\right)^{\frac{2}{n-1}} \sim\pi a^2
\left(\frac{V_0}{\Delta}\right)^{\frac{2}{n-1}} .
\end{equation}
(States with $ka \agt 1$ are activated due to the high
temperature).

In the most interesting case, when $n=3$, the cross-section
\begin{equation}\label{eq:sigma3}
\sigma \sim \pi a^2 \frac{V_0}{\Delta}
\end{equation}
can reach giant values $\sigma \sim 10^2 a^2$.
 Eq.~(\ref{eq:sigma-quasi}) shows that the effective
cross-section in the gas approximation (\emph{noninteracting}
quasiparticles) considered is inversely proportional to the band
width $\Delta$. The same dependence on $\Delta$ follows also from
the theories of Landesman \cite{Landesman75} and Yamashita
\cite{Yamashita74} for \emph{interacting} quasiparticles.  This
coincidence is evidently the explanation of the fact that the
dependence of the diffusion coefficient on the molar volume (``the
Gr\"uneisen parameter'') does not vary with decreasing
concentration \cite{AllenRichardsSchratter82} (see below).

The condition (\ref{eq:quasi-class}) can be satisfied at lower
temperatures as well if the following inequalities take place
\begin{equation}\label{eq:another case}
A(A/V_0)^{\frac{2}{n+1}}\ll T \ll \Delta
\end{equation}
Then the quasipartitle thermal energy may be evaluated from the
relation
 $\dx
 \hbar^2 k^2/2m^* \approx 3T/2.
 $
 With $m^* \approx \hbar^2/Aa^2$
  this yields $ka \sim (T/A)^{1/2}$, and the
cross-section (\ref{eq:sigma-quasi}) can be written in the form
\begin{equation}\label{eq:sigma(T)}
\sigma \approx \pi
a^2\left(\frac{V_0^2}{TA}\right)^{\frac{1}{n-1}}
{}_{\stackrel{-\!-\!\longrightarrow}{n=3}} \quad \pi
a^2\frac{V_0}{\sqrt{TA}}.
\end{equation}
It increases with decreasing temperature up to the value
\begin{equation}\label{eq:sigma_max}
\sigma_{max}\approx\pi a^{2}(V_0/A)^{\frac{2}{n-2}} \,\,
{}_{\stackrel{-\!-\!\longrightarrow}{n=3}} \,\, \pi a^{2}(V_0/A)^2
.
\end{equation}
The opposite situation which corresponds to the scattering of a
slow particle by a rapidly decreasing potential barrier $(n=6)$
can take place for vacancions at $T\ll\Delta$. Then the
cross-section does not depend on velocity, and hence, is
independent of temperature:
\begin{equation}\label{eq:sigma_slow}
\sigma \approx 3.65\pi a^2\left(\frac{V_0}{2A}\right)^{\frac12} .
\end{equation}
This expression fits well to that of Eq.~(\ref{eq:sigma_max}) at
$n=6$. If vacancion-vacancion scattering is under consideration,
then $V_0 \sim ms^2$,  $A = 1K$, and hence, $\sigma = 50 a^2$.

In the consideration above we did not take into account the role
of the anisotropy given by the factor $\Gamma(\mathbf{n})$ in the
potential (\ref{eq:fi_anisotrop}). It is obvious that such a
sign-variable function with zero averaged value can only reduce
the cross-section. According to Slusarev et al.
\cite{SlusarevFNT77} the anisotropy reduces the total
cross-section by a factor $0,105$ (cubic lattices), and $0,208$
(h.c.p. crystals). The reduction for the transport cross-section
is about 4-5 times. However, the exchange integral $A$ as well the
defecton band width $\Delta$ evaluated in the same work turned out
to be more than an order of magnitude less compared to the results
adopted now.

Situations where $\Delta\gg V_0$ are realized in vacancion
scattering by isotopic impurities. In such cases Born
approximation may be used and the problem can fully be solved in
an analytical form with an account of the exact form of the
function $\Gamma(\textbf{n})$ \cite{PhD,DiffusionJETP75}. It can
be shown that the scattering amplitude $F(\textbf{n})$ has the
same angular dependence as the interaction potential:
 $
 F(\textbf{n}) = \lambda\Gamma(\textbf{n})
  $.
  For a simple cubic
lattice $\lambda = (2a/15)(V_0/A)$, and therefore,
\begin{equation}\label{eq:sigma Born}
d\sigma = \frac{4}{225}a^2(V_0/A)^2\Gamma^2(\textbf{n})d\Omega ,
\end{equation}
\noindent where $\Gamma(\textbf{n}) = \sum_{i} n_i^4 - 3/5$.

One has to keep in mind, however, that the interaction potential
has the form (\ref{eq:fi_anisotrop}) only at large distances and
the effect of the scattering by the core needs for wide-band
vacancions a special consideration.

Consider briefly the role of the defecton statistics. If defectons
obey Fermi-Dirac statistics, then their quasimomenta has to be
near to the value of the Fermi-quasimomentum both before and after
collision. As far as the cross-section is proportional to the
final states number, it is proportional to temperature $T$.
Therefore,
\begin{equation}\label{eq:sigma Fermi}
\sigma_F = \sigma' T/ \epsilon_F .
\end{equation}
The quantity $\sigma '$ for the potential (\ref{eq:fi_general})
with $n = 3$ in quasiclassical approximation is of the form
\cite{PushkPSS76,Singapore}
\begin{equation}\label{eq:sigma-prim}
\sigma' = 2\pi^2\frac{V_0a}{\hbar v_F}
\end{equation}
\noindent where $v_F$ is the Fermi-velocity.  Therefore, the
cross-section depends not only on the ratio $V_0 /A$, but also on
the concentration:
\begin{equation}\label{eq:sigma_F}
\sigma_F \sim T\frac{V_0}{A^2x}, \quad T < Ax^{2/3} .
\end{equation}
The condition for the applicability of the  expressions
(\ref{eq:sigma-prim}) and (\ref{eq:sigma_F}) follows from
(\ref{eq:quasi-class}) where one must take into account, that
$k_{F}a\sim x^{1/3} $ :
\begin{equation}\label{eq:Fermi condition}
x > (A/V_0)^3 .
\end{equation}
It is worth noting also that the ratio $V_0 /A$ is strongly
dependent on pressure, and increases rapidly with decreasing molar
volume.

\section{Low Temperature Diffusion}

Let us now turn to the diffusion. As follows from the gas kinetics
theory the diffusion coefficient is equal to
\begin{equation}\label{eq:D=lv}
     D =  \frac{1}{3}lv  .
\end{equation}
\noindent where the mean free path is
\begin{equation}\label{eq:free path}
 l = a/ \sqrt{2} x \sigma
\end{equation}
This approximation is valid if the mean distance between
quasiparticles, $x^{-1/3}$, is much larger than the "radius" of
the cross-section  $\ R_0\sim\sqrt{\sigma}$, i.e.  if
 $\dx
x^{-1/3}\gg\sqrt{\sigma}
 $
 or (what is the same):
\begin{equation}\label{eq:condition1}
 x \ll 1/\sigma^{3/2}.
\end{equation}
The quasiparticle mean velocity $v$ depends, in general, on the
temperature. If $T\gg\Delta$ which is the case realized in
experiments on impurity diffusion in helium, then the levels till
the top of the energy band are populated. Sacco and Widom
\cite{SaccoWidom76} calculated the band-averaged value of the
velocity squared in a h.c.p. lattice and obtained (in our
notation)
\begin{equation}\label{eq:SaccoWidom}
  v_{SW} = ( <v^2>)^{1/2} = \frac{3}{\sqrt{2}}  \frac{a_0 A}{\hbar} =
    \frac{3}{2\sqrt{2}}s \frac{A}{\theta_p} \approx s \frac{A}{\theta_p}.
\end{equation}
In principle, this value can be used to estimate the order of
magnitude of the diffusion coefficient ({\ref{eq:D=lv}). However,
it is well known (see e.g. Ref.~\cite{Feynman}) that the correct
value of the velocity is closer to its highest value which can be
evaluated from the relation $\dx m^*v^2/2 = \Delta$:
\begin{equation}\label{eq:velocity}
v\approx s\frac{1}{\sqrt{2z}}\frac{\Delta}{\theta_p}\approx 2,45s
\frac{A}{\theta_p} \quad (z = 12).
\end{equation}
The diffusion coefficient is therefore (comp. to (\ref{eq:D})):
\begin{equation}\label{eq:D_0}
D_0=\frac{1}{6\sqrt z}\frac{as}{x\sigma}\frac{\Delta}{\theta_p}
\approx \frac{1}{\sqrt 3}\frac{as}{x\sigma}\frac{A}{\theta_p}
\quad (z = 12).
\end{equation}
Substituting here $\sigma$ from (\ref{eq:sigma3}) yields
\begin{equation}\label{eq:D_0I}
 D_0 =\frac{1}{6\pi\sqrt z}
 \frac{as}{x}\frac{\Delta^2}{V_0\theta_p}=
\frac{1}{3\pi\sqrt z}
 \frac{a^2}{\hbar x}\frac{\Delta^2}{V_0} .
\end{equation} In order for the
defectons to be good quasiparticles, their mean free path should
be larger than the lattice constant, i.e. the coherency condition
$x\sigma<1$ should be satisfied. This leads to the following
restriction on the defecton concentration:
\begin{equation}\label{eq:concentration condition}
x < x_0 \equiv \sigma^{-1} .
\end{equation}
As we showed in the previous section the cross-sections $\sigma$
can be extremely large. That is why the condition
(\ref{eq:concentration condition}) is weaker than
(\ref{eq:condition1}). Hence, the gas-kinetics formula
(\ref{eq:D_0}) is valid for concentrations
\begin{equation}\label{eq:x-condition}
x < \sigma^{-3/2} < \sigma^{-1} = x_0  .
\end{equation}
The concentration $x_0$ turns out to be smaller than the critical
concentration $x_c$ (\ref{eq:x_c-value}). This means that
\emph{the localization process may be considered in terms of
quasiparticles}. Making use of the results obtained we may
distinguish the following concentration regions depending on the
powers of the small parameter $\xi =\Delta/V_0$ :
\medskip

     I. $ \dx  x < \xi^{3/n-1}\leq 10^{-3}$  --- gas approximation

    II. $ \xi^{3/n-1} < x < (\xi/n)^{3/n+1}\sim 10^{-2}$ ---
interacting quasiparticles

   III. $ (\xi/n)^{3/n+1} < x $   --- localized defectons

\noindent(The numerical values are calculated for $n = 3$)
\medskip

Let us consider now the diffusion in the second region.  In this
case the mean free path $l$ is controlled by the \emph{scattering
on the defecton band boundaries} (Fig.~\ref{fig:1}.), i.e.  by the
condition $l\mid\bigtriangledown\varphi\mid \approx\Delta$.  This
means that the variation of the defecton energy along a distance
$l$ should not exceed the band width $\Delta$. The corresponding
diffusion coefficient in this region is therefore of the form
\cite{LandesmanWinterLT13,Huang-Guyer75,AndreevUFN76}:
\begin{equation}\label{eq:D-interact}
D \sim \frac{\Delta^{2}}{V_0 x^{(n+1)/3}}
\end{equation}

It is worth noting that the cross-sections both in region II, and
in region I depend on $\Delta$.  This dependence is important not
only for choosing suitable defect concentration and perfection of
the crystal samples.  The knowledge of the diffusion coefficient
as a function of $\Delta$ allows to find the dependence of $D$ on
the molar volume $V_m$.  Since this dependence is strong and
different in different concentration regions, one may use it to
detect a change of the scattering mechanism.

At temperatures lower than the degeneracy temperature $T_0$,
diffusion depends on defecton statistics  \cite{PushkPSS76}.  If
the quasiparticles obey Fermi-statistics, then the gas
approximation yields
\begin{equation}\label{eq:D_F}
 D_F = \frac{1}{3}l_Fv_F, \quad {\rm where} \quad v_F  =
(3\pi^2)^{1/3}\frac{Aa}{\hbar} x^{1/3}
\end{equation}
\noindent and the mean free path is
$$
 l_F = \frac{a}{\sqrt{2}\sigma x_{eff}} .
$$
Since only defectons near the Fermi-surface take part, the
``active concentration'' is equal to
$$
x_{eff} = x\frac{T}{\epsilon_F}
$$
\noindent and the effective cross-section is given by
Eqs.~(\ref{eq:sigma Fermi})--(\ref{eq:sigma_F}).  As a result, the
mean free path and the diffusion coefficient are multiplied by an
additional factor $(\epsilon_F /T)^2$.  This yields
\begin{equation}\label{eq:D_Fermi}
D_F = \frac{1}{3}\frac{av_F}{x\sigma_0}
\left(\frac{\epsilon_F}{T}\right)^2
\end{equation}

Taking into account (\ref{eq:sigma-prim}) one may rewrite
Eq.~(\ref{eq:D_Fermi}) in the form
\begin{equation}\label{eq:Fermi D}
D_F = \frac{3\pi^2}{16}as\frac{A^4}{\theta_p V_0 T^2}x .
\end{equation}
Of course, the increase of $l_F$ and $D_F$ with decreasing
temperature is limited by the finite size of the sample, or by
scattering on other crystal defects.

In the case of Bose-defectons the diffusion coefficient may  be
estimated taking into account that the ``actual'' concentration
depends only on temperature:
$$
x \approx\frac{1}{6}\left(\frac{T}{A}\right)^{3/2}
$$
\noindent and hence, the mean free path is
$$
l \approx \frac{6a}{\sqrt{2} \sigma}\left(\frac{A}{T}\right)^{3/2} .
$$

The mean velocity is
$$
v \approx 1,75\frac{a}{\hbar}(AT)^{1/2} .
$$
The diffusion coefficient is therefore
$$
D \approx 1,25 \frac{as}{\sigma}\frac{A^2}{\theta_p T} .
$$
If the mean free path turns out to be larger than the sample size
$L$, then
$$
D \sim \frac{La}{\hbar}(AT)^{1/2} .
$$

One has to keep in mind, however, that those effects can be
violated by a superfluid current, which can take place even at
small chemical potential gradient.

Apparently, the diffusion at $T < T_0$ is not observable in helium
due to the rapid decreasing of vacancion concentration (except of
zero-point vacancions, not found yet) and the phase separation in
the case of impuritons. To observe it one may have to look for
another quantum system.

 \section{Phonon Controlled Diffusion}

At absolute zero temperature a defecton moves through a crystal
like a free particle without dissipation.  In view of its low
velocity compared with the sound one, the gas of zero-point
phonons has time to adjust itself to the defecton which moves
together with an adiabatically adapted phonon ``cloud'' that
strongly influences only its effective mass.  Note, that
null-phonons do not change the periodicity of the lattice.

With rising temperature, scattering of thermal phonons by the
defecton sets in. Since the defecton mass is large and its
lifetime is much longer that the relaxation time of the phonon gas
the cross section for scattering by a defecton can be replaced in
zeroth approximation with respect to the small parameter $(v/s)^2$
by the cross section for scattering by an immobile defect. As it
has been shown by I.M.Lifshitz \cite{Lifshitz1948} (see also
\cite{LifshitzKaganovUFN,LifshitzAzbelKaganov}) such a problem can
be solved in an analytical form \cite{JETP70,Singapore,Nauka}.

The differential cross-section for the scattering of a phonon of
the $j$-th mode with polarization $\nu$ as a result of which a
phonon of the $i$-th mode with a polarization $\nu'$ is produced
is given by the expression \cite{JETP70}:
\begin{eqnarray}\label{eq:d sigma}
d\sigma^{\nu \nu'}_{j\mathbf{k},i\mathbf{k'}}&&=
\left(\frac{V}{4\pi m s_i^2}\right)^2 \left(q^\nu_i\right)^2
\nonumber\\
&&\times \left\{\sum_{\mu\mu'}q_i^\mu q_j^{\mu'} \sum_{R R'}
\mathbf{kR}\Lambda_{RR'}^{\mu\mu'} \mathbf{k_i'R'}\right\}^2 d
\Omega'
\end{eqnarray}
where $V$ is the lattice cell volume, $s_i$ is the sound velocity
of the corresponding phonon branch, $q_j^\nu$ is the polarization
vector and $d \Omega'=2\pi \sin\alpha d\alpha$, $\alpha$ being the
angle between the directions of the incident and scattered
phonons. The matrix $\Lambda^{\mu\nu}_\mathbf{R R'}(\mathbf{r})$
describes the change of the elasticity matrix caused by the defect
(situated in the lattice site \textbf{r}). Averaging over the
modes and polarizations of the incident phonons and summing over
the modes and polarizations of the scattered ones yield

\begin{equation}\label{eq:sigma_kk}
d\sigma_{kk'} = \left(\frac{V}{2\pi}
\right)^{2}k^{4}\Sigma_{\mathbf{n n'}}d\Omega'
\end{equation}
\noindent where $\mathbf{n}= \mathbf{k}/k$, $\mathbf{n'}=
\mathbf{k'}/k$,
 $$
 \Sigma_\mathbf{nn'}= \frac{V}{12 m^2
 s^4}\sum_{\mu\nu}\left\{\sum_\mathbf{RR'}\mathbf{nR\Lambda_\mathbf{RR'}
 ^{\mu\nu}}\mathbf{n'R'}\right\}^2 .
 $$
 Note, that the averaged value of the sound velocity
 $ \dx
 s= (s_l^{-4} + 2 s_\perp ^{-4})^{-1/4}
 $
 does not coincide with the Debye velocity, even if does not
 differ significantly.

 Any further application of the formula (\ref{eq:sigma_kk})
requires some model assumptions regarding the matrix components
 $
\Lambda_{\mathbf{RR'}}^{\mu\nu}
 $.
 They have to obey, however, some relations that follow from
the behaviour of the forces in a crystal lattice \cite{Maradudin}.
In the nearest neighbour approximation this allows to express all
matrix elements by one of them (e.g. $\Lambda_{00}$)
\cite{JETP70}. In case of a vacancy in a simple cubic lattice
$\Lambda_{00}= -6 ms^2/a^2$ and no free parameters are required.
In this case $V = a^3$, $\Sigma_\mathbf{nn'} = (\mathbf{nn'})^2$
and
\begin{equation}\label{eq:sigma_k^4}
d\sigma_\mathbf{kk'} = \left( \frac{a}{2\pi}
\right)^2(ka)^4(\mathbf{nn'})^2 d\Omega' .
\end{equation}
In case of impurity, this quantity should be multiplied by a
factor
 $
\sigma_0 =(\Lambda^{impurity}_{00}/\Lambda^{vacancy}_{00})^2
 $
Note, that during all the consideration we have not supposed the
perturbation caused by the defect small. This is the advantage of
the Lifshitz method used. Let us call attention to the following
peculiarities when considering isotope defects. In a classical
crystal isotope defects interact with the host atoms in
approximately the same way and the main difference consists in
their mass. This leads to a correction factor $\dx \sigma_0 = (1-
m_{host}/m_{isotope})^2$. For $^3$He impurities in solid $^4$He
$\sigma_0=1/9$. In case of quantum crystals isotopes differ not
only in mass but also in their zero-point vibrations affecting in
this way the local values of the elasticity module
\cite{Singapore}. An attempt to estimate this effect was made by
Slusarev et al. \cite{SlusarevFNT78}. They argued that this effect
could be larger than that of the mass-difference and that we had
not taken it into account. The later statement is not acceptable
as seen by the consideration above. In our short letter
\cite{JETPLett74} the principal value of the cross-section (given
by the mass-difference) was used in order to avoid model
considerations. We shall not discuss here the model assumptions in
\cite{SlusarevFNT78} in detail. However, the correction proposed
is given by the quantity $\tilde\alpha\approx \Theta_3/\Theta -1
\approx 0.23$ in the expression (cf. Eq.(15) in
Ref.\cite{SlusarevFNT78})
\begin{equation}\label{eq:SlusarevG}
 G_1/4 = \alpha^2 +{\tilde\alpha}^2  - \frac{1}{2}\alpha{\tilde\alpha}
 \approx
0.11 +0.0145
\end{equation}
with $\alpha = \sigma_0 = 1/3$. Obviously, the correction 0.0145
could not be considered as 'larger' than 0.11. In our opinion next
terms in Eq.(14) in \cite{SlusarevFNT78} of the order of
$G_1/2^{12}\sim 2\times 10^{-4}G_1$ are far from the model
accuracy used.

 The diffusion coefficient may be
calculated solving the Boltzmann equation, which is of the type of
Fokker-Plank equation \cite{Singapore,Nauka}. However, the
specific features of the defecton motion in the phonon gas allows
to solve the problem without resorting to the kinetic equation.
We follow here the method proposed in
\cite{JETP70,DiffusionJETP75}.

Since the defecton velocity is much smaller than the sound one the
main contribution is due to two-phonon processes. The energy and
quasimomentum conservation laws for an individual collision are:
\begin{equation}\label{eq:E-conserv}
\epsilon(\textbf{p}) + \hbar sk = \epsilon(\mathbf{p'}) + \hbar
sk'
\end{equation}
\begin{equation}\label{eq:p-conserv}
\textbf{p} + \textbf{k} = \textbf{p}'+\textbf{k}'+2\pi \textbf{b}
\end{equation}
\noindent where $\textbf{k}$ and $\textbf{k}'$ are the quasiwave
vectors of the incident and scattered phonons respectively,
$\epsilon(\textbf{p})$ -- the defecton dispersion law, and
\textbf{b} is an arbitrary vector of the reciprocal lattice.  The
linear dispersion law for the phonons is used since the
temperature is much smaller than the Debye one.  The phonon gas
can be considered as in equilibrium, and hence the mean value of
the phonon quasimomentum is $\hbar k\propto T/s$. Only phonons
with quasiwave vectors near the centre of the Brillouin zone take
part in collisions, and Umklapp processes may be neglected.  The
temperature, in general, can be higher or lower than the defecton
band width $\Delta$. In both cases the relative variation of the
defecton energy and quasimomentum in a collision are small. If
$\Delta\ll T\ll\theta_p$ then  Eqs.~(\ref{eq:E-conserv}) and
(\ref{eq:p-conserv}) yield
$$
\delta\epsilon = \frac{\partial\epsilon}{\partial
\textbf{p}}\delta \textbf{p}  = \hbar
\textbf{v}(\textbf{p}-\textbf{p}') = \hbar \textbf{v(k-k')} \leq
2v\hbar k\propto 2vT/s
$$
\noindent and hence,
\begin{equation}\label{eq:E-change}
\delta\epsilon/ \Delta \leq T/ \theta_p \ll 1
\end{equation}

In the same way
\begin{equation}\label{eq:p-change}
\frac{\mid \textbf{p - p}' \mid}{p} = \frac{\mid \textbf{k  - k}'
\mid}{p}\leq\frac{2T}{\hbar sp}\sim\frac{T}{pa\theta_p} \ll 1,
\quad v\propto a \Delta /\hbar
\end{equation}
In the opposite case, when $T < A < \theta_p $, the defecton
dispersion law may be replaced by a quadratic one and the
effective mass approximation can be used:
$$
\epsilon(p) =\frac{\hbar^2 p^2}{2m^{\star}}
$$
Then the change of the phonon wave vector is
\begin{equation}\label{eq:k-change}
\chi = k' - k= \frac{k}{p_0} \textbf{p(n'-n)}
-\frac{k^2}{p_0}(1-\textbf{nn}')
\end{equation}
\noindent where $ p_0 = (m^{\star}s)/\hbar = 2\theta_p/(Aa)$
and $\textbf{n} = \textbf{k}/k$, $\textbf{n}' = \textbf{k}'/k$ ---
the directions of the incident and scattered phonons. The wave
vector transferred is
\begin{equation}\label{eq:q}
\textbf{q} = k(\textbf{n}-\textbf{n}') + \chi \textbf{n}'.
\end{equation}
The relative variations of energy and  quasimomentum  are,
therefore, small in this case too
$$
\frac{\delta E}{\epsilon(p)}\propto \frac{k}{p}\propto
\frac{(AT)^{1/2}}{\theta_p}\ll  1,
\quad \frac{q}{p}\propto\frac{k}{p}\ll 1 .
$$
As for the phonon quasimomentum, its direction changes greatly in
each collision, but its magnitude remains practically constant.
The situation thus recalls the motion of a heavy particle in a gas
of light particles.  The difference lies in the dispersion laws
and in the different statistics obeyed by the defectons and
phonons.  It is important also that the width of the defecton
energy band is relatively small.

Let us find first the diffusion coefficient in momentum space.
Obviously, the change of the square of the wave vector per unit
time is given by the integral
\begin{equation}\label{eq:q^2}
\langle q^2\rangle = \frac{3s}{(2\pi)^3 a^5} \int{d^3
kd\sigma_\textbf{kk'} n(k)(\textbf{k}-\textbf{k}')^2}
\end{equation}

\noindent where $n(k) = \{ \exp (\hbar sk/T)-1\}^{-1}$ is the
phonon distribution function, and $d\sigma_{\textbf{kk}'}$ is the
differential cross section (\ref{eq:sigma_kk}).  At low
temperatures $(T\ll\theta_p)$ the integral in Eq.~(\ref{eq:q^2})
may be expressed in terms of the Riemann $\zeta$-function:
\begin{equation}\label{eq:<q^2>}
\langle q^2\rangle =  \alpha\frac{\sigma_0
s}{a^3}\left(\frac{T}{\theta_p}\right)^9, \quad \alpha =
\frac{\pi^5\zeta(9)}{240\zeta(8)} \approx 1,27 .
\end{equation}

To calculate the diffusion coefficient in coordinate space we
proceed as follows.  We define the free path time of the defecton
as a time during which the transferred momentum squared becomes of
the order of the square of the initial particle momentum:
\begin{equation}\label{eq:tau}
\tau = \tau_0 \frac{1}{\alpha\sigma_0}
\frac{\overline{\epsilon(p)}}{\theta_p}
\left(\frac{\theta_p}{T}\right)^9 , \quad \tau_0 = \hbar/A
\end{equation}

Then the mean free path is the path traversed during the time
$\tau$ :
\begin{equation}\label{eq:l}
l = \frac{\sqrt2}{\alpha} \frac{a}{\sigma_0}
\frac{\overline{\epsilon(p)}^{3/2}}{\theta_p\sqrt{A}}
\left(\frac{\theta_p}{T}\right)^9 ,
\end{equation}
\noindent and the diffusion coefficient in the phonon gas is
\begin{equation}\label{eq:D-T^9}
D_T =
\frac{as}{3\alpha\sigma_0}\left(\frac{\overline{\epsilon(p)}}
{\theta_p} \right)^ 2 \left(\frac{\theta_p}{T}\right)^9 .
\end{equation}

The presence of the ninth power the temperature has a lucid
physical meaning -- three degrees are connected with the number of
phonons,  four with the scattering cross section, and two with the
``ineffectiveness'' of the collisions (when a heavy particle is
scattered by light ones).

It is instructive to note once again that in all the equations the
characteristic temperature parameter is not the Debye temperature
$\Theta$ as assumed in
\cite{AL69,GriJETP73,AndreevUFN76,Esselson-book82} but $\theta_p
\approx \Theta/8$.  The effect of this is that the kinetic
characteristics evaluated in the works mentioned differ from our
results by a great factor of the order of $10^8$.

The mean value $\overline{\epsilon(p)}$ depends both on the
temperature region and on the quasiparticle statistics. For
Fermi-Dirac statistics,  $\overline{\epsilon(p)}= \epsilon_F
\propto \Delta x^{2/3}$ , and

\begin{equation}\label{eq:D_F(T)}
D_F\propto\frac{as}{\sigma_0} x^{4/3}
\left(\frac{\Delta}{\theta_p}\right)^2 \left(\frac{\theta_p}{T}
\right)^9 .
\end{equation}
In the case of Bose-Einstein statistics:

\begin{equation}\label{eq:D_Bose}
D_B\propto\frac{as}{\sigma_0}\left(\frac{\theta_p}{T}\right)^7 .
\end{equation}
Since the degeneracy temperature $T_0\propto Ax^{2/3}$ is
extremely  small, the inequality $T_0\ll T$ is practically
satisfied in the whole experimentally accessible region.  If at
the same time $T < \Delta$, then Boltzmann statistics applies and
the relation $\epsilon(p) = 3T/2$ may be used to estimate the
averages over the temperature.  In this case we obtain for the
free path time and for the diffusion coefficient, respectively,

\begin{equation}\label{eq:tau_Bose}
\tau_B\approx   1,18\frac{\tau_0}{\sigma_0}
\left(\frac{\theta_p}{T} \right)^8 , \quad D_B\approx
0.59\frac{as}{\sigma_0} \left(\frac{\theta_p}{T} \right)^7 .
\end{equation}

From the last expression and the Einstein relation we easily
obtain the mobility $b$ of the defecton in the phonon gas
\begin{equation}\label{eq:Bose-mobility}
b=\frac{D_B}{T} = 1.18\frac{a^2}{\hbar\sigma_0}
\left(\frac{\theta_p}{T} \right)^8 .
\end{equation}
Naturally, the same value of the mobility can be obtained by
direct calculation of the force experienced by a quasiparticle
moving in the gas.  Indeed, in the system in which the defecton is
at rest, the phonon distribution function takes the form
$n(\textbf{k}) = \tilde{n}(\varepsilon - \hbar \textbf{kv})$, and
the force acting on the defecton can be expressed as the change of
momentum per unit time:
$$
\textbf{F}= \frac{3s\hbar}{(2\pi)^3}\int d^3
kd\sigma_{\textbf{kk}'}\tilde{n}(k)k(\textbf{n}-\textbf{n}')=
b^{-1}\textbf{v} .
$$
In first order in the small ratio $v/s\ll 1$, the mobility
calculated  in this manner coincides with that of
Eq.~(\ref{eq:Bose-mobility}).

If $T >\Delta$, then the band is filled uniformly and
$\overline{\epsilon(p)} = \Delta$.  In this region which is the
most interesting one from the experimental point of view we have

\begin{equation}\label{eq:T^9}
\tau =\frac{\tau_0}{\alpha\sigma_0} \frac{\Delta}{\theta_p}
\left(\frac{\theta_p}{T} \right)^9, \ D_T =
\frac{as}{3\alpha\sigma_0} \left( \frac{\Delta}{\theta_p}
\right)^2 \left(\frac{\theta_p}{T} \right)^9 .
\end{equation} 

It follows from the deduction made that the results obtained are
valid as far as the mean free path $l$ is much longer than the
lattice constant $(l\gg a)$. This condition leads to the
inequality $T < T_k$, where

\begin{equation}\label{eq:T_k}
T_k = \theta_p \left( \frac{\sqrt{2z}}{\alpha\sigma_0}
\frac{\Delta}{\theta_p} \right)^{1/9} .
\end{equation} 

At temperatures $T > T_k$ the mean free path becomes smaller than
the interatomic distances, and the free path time $\tau$ gets
shorter than the delocalization time (\ref{eq:tau_0}).  As a
result, the defecton spends a greater part of the time within the
cell, and only rarely does it execute individual transitions
(tunnel or activation) to a neighbouring equivalent position ---
the so called \emph{ phonon stimulated localization} takes place.
At $T = T_k$ the diffusion coefficient is
\begin{equation}\label{eq:D(T_k)}
D \propto \frac{a^2 \Delta}{\hbar}
\end{equation}
Its behaviour at higher temperatures calls for a special analysis.

It has been reported in \cite{KaganKlinger}  that in a case of
light interstitials the law $D\sim T^{-9}$ is valid up to the
Debye temperature, independently of the small mean free path.
Unfortunately, the melting point of solid helium is much lower the
Debye temperature, and hence this prediction cannot be tested
experimentally.  On the other hand the consideration made in
\cite{KaganKlinger} does not take into account that the activation
mechanism may become essential far below the Debye temperature
$\Theta$.

\section{Comparison with Experiment. Determination of the
 Impuriton Characteristics.}

Let us summarize the results obtained and compare them with the
available experimental data on diffusion of He impurities in solid
He.

The most important processes which determine quantum diffusion in
helium are:  i)~scattering by defects, leading to a temperature
independent diffusion, and ii)~defecton-phonon scattering which
determine the temperature dependence. In the most interesting case
when $T\gg\Delta$ the diffusion coefficient is given by
Eq.~(\ref{eq:T^9}). As far as the defecton-defecton and
defecton-phonon scattering are independent, the total diffusion
coefficient may be obtained using the Matthiessen rule.  This
yields \cite{JETPLett74}:
\begin{equation}\label{eq:D^-1}
D^{-1} = D_0^{-1} + D_T^{-1} = \frac{\theta_p}{as\Delta} \left[
2\sqrt{z}x\sigma + 3\alpha\sigma_0\frac{\theta_p}{\Delta} \left(
\frac{T}{\theta_p} \right)^9 \right] .
\end{equation}
The transition from the temperature regime (\ref{eq:T^9}) to the
constant value (\ref{eq:D_0}) takes place at the temperature
\begin{equation}\label{eq:T*}
T^{\star} = T_k l^{-1/9}
\end{equation}
\noindent where $T_k$ is given by the expression (\ref{eq:T_k}).
In that way the temperature interval may be divided into three
parts. The region $\Delta <T < T^{\star}$ corresponds to
impuriton-impuriton scattering while scattering by phonons
prevails at $T^{\star} < T < T_k$.  At $T > T_k$ (but below the
melting point) the classical activation diffusion comes into
effect. The diffusion coefficient behaviour is schematically shown
in Fig.~\ref{fig:SchematicD}.  The diffusion coefficient first
decreases exponentially with decreasing temperature to $T_k$, then
increases rapidly $(\sim T^{-9})$, and when $T < T^{\star}$ goes
to a plateau which depends on the concentration (to be more
precise, on the mean free path $l \sim 1/(x\sigma)$). If
$x\sigma\rightarrow 1$, then the mean free path $l\rightarrow 1$,
and $T^{\star}\rightarrow T_k$. It is seen, therefore, that the
observation of the temperature dependence of the coherent quantum
diffusion is possible as far as the coherency condition $l\gg 1$
is fulfilled.

Let us consider for definiteness the case of h.c.p.  $^4$He
crystal with molar volume $V_m \approx 21\ \mathrm{cm}^{3}$.  Then
the values of the quantities included in the expression
(\ref{eq:D^-1}) which can be measured independently are as follows
\begin{equation} \label{eq:values}
\theta = 26\ \mathrm{K}, \ \theta_p = 10/3 \mathrm{K},\ a = 3,27
\AA, \ s = 320\ \mathrm{m/s} .
\end{equation}
The quantity $\sigma_0$ calculated with an account of the mass
difference of helium isotopes is equal to \cite{JETPLett74}
$\sigma_0 = ((m_4 - m_3) /m_3)^2 = 1/9$. We shall limit ourselves
with this value because an account of the change of the elastic
moduli \cite{SlusarevFNT78} would be based on model considerations
and would finally give a correction small compared to the
experimental error.

Substituting (\ref{eq:values}) into (\ref{eq:D^-1}) yields the
following expression for the dependence of the diffusion
coefficient on the temperature and concentration.
\begin{equation}\label{eq:D^-1number}
D^{-1} = 5.6 \times 10^4\frac{x\sigma}{\Delta} +
\frac{0.1}{\Delta^2} T^9 .
\end{equation}
The expressions for characteristic temperatures then take the
form:
\begin{equation} \label{eq:T*_k}
T_k = 4.2\Delta^{1/9}, \quad T^{\star} = T_k l^{-1/9} \approx
T_k(x\sigma)^{1/9}
\end{equation}

The second term in (\ref{eq:D^-1number}) corresponding to the
diffusion law (\ref{eq:T^9}) allows to determine the band width
$\Delta$ \emph{from the temperature dependence only }. Having the
value of $\Delta$ one can calculate the cross-section $\sigma$
from the first addend.

In order to study the dependence on the molar volume $V_m$ let us
write the expression (\ref{eq:D^-1}) in the form:
\begin{equation}\label{eq:D^-1short}
D^{-1} = B^{-1} x + G^{-1} T^{9}
\end{equation}
 where
\begin{equation} \label{eq:B&G}
B=D_0x, \quad G = 2.4 as \Delta^2 \theta_p^7 =
4.8\frac{a^2}{\hbar}\Delta^2 \theta_p ^8.
\end{equation}
The values of the coefficients $B$ and $G$ at different molar
volumes and concentrations are given in Table 1.  As far as the
dependence of $\theta_p$ on $V_m$ is known (the Gr\"uneisen
parameter $\gamma$ is the same as for the Debye temperature).
Eq.~(\ref{eq:D}) enables us to find the "Gr\"uneisen parameter for
the band width" $\gamma_{\Delta}$ :

\begin{equation}\label{eq:gamma_delta}
\gamma_{\Delta} = d (\ln \Delta) / d (\ln V_m) = 4\gamma -
\frac{1}{3} + \frac{1}{2}\gamma_G
\end{equation}
\noindent where

$$
\gamma_G = d (\ln G) / d (\ln V_m) .
$$
Substituting here $\gamma = 2,6$ and calculating $\gamma_G\approx
21$ from the experimental data given in Table~1 yield
 $\dx
\gamma_{\Delta} \approx 21 .
 $

\begin{table}\caption{\label{tab:table1}
 Quantities $B$, $G$ (\ref{eq:D^-1short}) and $\sigma$ as functions
of the molar volume $V_{m}$ and concentration $x$.}
\begin{ruledtabular}
\begin{tabular}{|l|l|c|c|c|r|} \hline
\ \ $x$  & \ \ $V_{m}$  & $B.10^{11}$ & $G.10^{-8}$ & $\sigma$ &
source \\
\ \ $\%$ & \ \ $cm^{3}$ & $cm^{2}/s$  & $cm^{2}s^{-1}K^{-9}$ &
 & \\   \hline
0.05   & 20.23 & 0.35  &  4.8 & 173 & \cite{AllenRichardsSchratter82} \\
0.05   & 20.42 & 0.45  &  7.7 & 162 & \cite{AllenRichardsSchratter82} \\
0.05   & 20.62 & 0.9   &  4.5 & 104 & \cite{AllenRichardsSchratter82} \\
0.05   & 20.84 & 1.6   & 10   & 71  & \cite{AllenRichardsSchratter82} \\
0.05   & 20.98 & 2.8   & 10   & 48  & \cite{AllenRichardsSchratter82} \\
0.01   & 20.95 & 2.32  &      &     & \cite{SchratterAllen84} \\
0.01   & 21    & 2.8   &  10  &     & \cite{AllenRichardsSchratter82} \\
0.02   & 21    & 2.8   &  10  &     & \cite{AllenRichardsSchratter82} \\
0.0242 & 20.95 & 2.35  &      &     & \cite{SchratterAllen84} \\
0.0499 & 20.95 & 1.9   &      &     & \cite{SchratterAllen84} \\
0.05   & 21    &       &  24  &     & \cite{EsselsonNT20} \\
0.006  & 20.95 & 1.0   &  24  &     & \cite{EsselsonJETP78} \\
0.12   & 21    &       &  24  &     & \cite{EsselsonNT20}\\
0.25   & 21    & 0.81  &      &     & \cite{GriJETP74}\\
0.01   & 20.70 & 1.1   &      &     & \cite{SchratterAllen84}\\
0.242  & 20.70 & 1.2   &      &     & \cite{SchratterAllen84}\\
0.0499 & 20.70 & 0.9   &      &     & \cite{SchratterAllen84}\\
\hline
\end{tabular}
\end{ruledtabular}
\end{table}

This value is close to that for the Gr\"uneisen parameter for the
exchange integral in $\mathrm{^3 He}$~  $(\gamma_j\approx18)$.
Unfortunately, the experiment is not precise enough to lead to
unique conclusions.

Let us turn now to Eq.~(\ref{eq:D_0I}) and write the Gr\"uneisen
parameter in the form
\begin{equation}\label{eq:Gamma_D}
\Gamma = \partial (\ln D) / \partial (\ln V_m) = 2\gamma_{\Delta}
- \gamma_{v_{0}} + \frac{2}{3} \approx 42\frac{2}{3} -
\gamma_{v_{0}} .
\end{equation}

The experimental values of $\Gamma$ differs significantly even
within the publications of one and the same group. In
\cite{GriJETP74} the value
\begin{equation}\label{eq:Gamma47}
\Gamma = 47 \pm 11 ,
\end{equation}
\noindent has been obtained while Allen, Richards and Schratter
\cite{AllenRichardsSchratter78,AllenRichardsSchratter82,SchratterAllen84}
reported
\begin{equation}\label{eq:Gamma57}
\Gamma = 57 \pm 12 .
\end{equation}
A conclusion may be made from a comparison of (70) and (71) that
the interaction between impurities (characterized by the quantity
$V_0$) depends on the molar volume relatively small:
$\gamma_v\approx 5$.  Such a dependence can be obtained assuming
that $V_0\sim K_T a^3$, where $K_T$ is the isothermal modulus of
elasticity \cite{EsselsonNT20}.  On the other hand a comparison of
(70) and (72) leads to a strong dependence:  $\gamma_v\approx 16$.

Let us show now the order of calculation of defecton
characteristics  on the bases of the data in Table~1.  Estimation
will be made for $V_m = 21~\mathrm{cm}^3$, $x = 0.05 \%$, $B =
2.8\times 10^{-11}$ \cite{AllenRichardsSchratter82}. Comparing the
values for $G$ with Eq.~(\ref{eq:D^-1number}) yields the band
width $\Delta = 10^{-4}\ \mathrm{K}$. Substituting the result
obtained into the expression for $B$ (the first term in
(\ref{eq:D^-1number})) at given concentration one obtains the
cross-section $\sigma= 10^4 \Delta /(5,6 B)= 64 a^2$.  Then $V_0$
may be obtained from (\ref{eq:sigma3}): $V_0 \sim \sigma\Delta
\sim 0,65\times 10^{-2}\ \mathrm{K}$ (Eq.~(\ref{eq:sigma}) gives
slightly larger value of the same order.) Making use of
(\ref{eq:free path}) one may calculate the mean free path $l
=a/(\sqrt{2} x\sigma) \approx 23~a $ and verify the coherency
condition $l\gg a$. Finally, temperatures $T_k$ and $T^{\star}$
are calculated and compared with the experimental curve $D(T)$.
This procedure yields:
$$
\Delta\approx 10^{-4} \mathrm{K}, \quad \sigma\sim 64~a^2, \quad
V_0\sim10^{-2} \mathrm{K}
$$
\begin{equation}\label{eq:TkT*}
 T_k = 1.2~\mathrm{K},
\quad T^{\star} = 0.7(10^4 x)^{1/9} \mathrm{K}.
\end{equation}
If the value $G = 2.4\times 10^7$ reported in
Refs.~\onlinecite{EsselsonNT20,EsselsonJETP78} is used, then $\dx
\Delta \approx 1.5 \times 10^{-4}~ \mathrm{K}$. Both values are
within the experimental error.

In regard to the concentration regions one obtains that the gas
approximation is valid at concentrations
\begin{equation}\label{eq:x_I}
x < \sigma^{-3/2} \approx 10^{-3} ,
\end{equation}
and the region of interacting quasiparticles is given by the
inequalities

\begin{equation}\label{eq:x_II}
  10^{-3}  < x < 10^{-2} .
\end{equation}

A comparison of theoretical and experimental results
\cite{EsselsonNT20,AllenRichardsSchratter82,EsselsonJETP78} is
shown in Figs.~\ref{fig:T^9}, \ref{fig:D_0(x)}, \ref{fig:D(T)},
\ref{fig:V20_98}, \ref{fig:V20_84}, \ref{fig:V20_62}.
Fig.~\ref{fig:T^9} is an experimental corroboration of the ninth
degree in the diffusion law (\ref{eq:T^9}), Fig.~\ref{fig:D_0(x)}
shows the concentration dependence of the diffusion coefficient in
the region of the low-temperature plateau. The temperature
dependencies at different concentrations and molar volumes are
shown in Fig.~\ref{fig:D(T)} and Figs.~\ref{fig:V20_98},
\ref{fig:V20_84}, \ref{fig:V20_62}.

In order to avoid any misunderstanding it is worth nothing that an
interpolation formula analogous to (\ref{eq:D^-1}) has been used
also in \cite{GriJLTP73,GriJETP74,GriFNT75,QuantumLiquidsBook}:
\begin{equation}\label{eq:D_wrong}
D^{-1} = \frac{\hbar}{Aa^2} \left[x+\frac{\theta}{A}
\left(\frac{T}{\Theta} \right)^9 \right] .
\end{equation}

This expression contains two essential differences compared to
(\ref{eq:D^-1}). The first one is the presence of the Debye
temperature $\Theta\approx 8\theta_p$ instead of $\theta_p$. The
second one is that (\ref{eq:D_wrong}) does not contain the
cross-section $\sigma$ which role has already been considered. The
same error is present also in
\cite{RichardsPopeWidom72,WidomRichards72}. That is why the values
of the band width obtained in these works turned out to be smaller
by more than 2 orders of magnitude.

In order to describe the temperature dependence in the third
region in Fig.~\ref{fig:SchematicD} one has to add the classical
expression
\begin{equation}\label{eq:Dclassic}
    D_v = \frac{a^2}{6}\omega_v \exp\{-\Phi_v/T\}
\end{equation}
where the subscript $v$ refers to vacancies because the diffusion
is dominated by the vacancy-impurity exchange. Typical values of
the exchange frequency is $\omega_v \sim 10^8 \div 10^9\
\mathrm{s^{-1}}\sim 10^{-2}~\mathrm{K}$. The final form of the
diffusion coefficient is therefore:
\begin{eqnarray}\label{eq:D-Final}
 \!\!\!\!   D_{tot}\! &=& D + D_v \nonumber\\
   \! &=& \left\{ C_1 \frac{x\sigma}{\Delta} +
 C_2 \frac{\theta_p^2}{\Delta^2}
\left(\frac{T}{\theta_p}\right)^9\right\}^{-1}\! + D^0_v
e^{-\Phi_v/T}
\end{eqnarray}
where $D$ is given by eq.~(\ref{eq:D^-1}), and
\begin{equation}\label{eq:C_1}
C_1 = \frac{\hbar\sqrt{z}}{k_B a^2}\approx (2.5 \pm 0.5) \times
10^4~ \mathrm{K.s/cm^2},
\end{equation}
\begin{equation}\label{eq:C_2}
 C_2 = \frac{3\alpha\sigma_0}{as}= (440\pm 15)~\mathrm{cm^{-2}s},
\end{equation}
\begin{equation}\label{eq:D^0_v}
  D^0_v =
\frac{1}{6}a^2 \omega_v = (3\div 10)\times 10^{-7}~
\mathrm{cm^2/s}
\end{equation}
with $k_B$ -- the Boltzmann constant. The quantities $\theta_p$
and $\Delta$ depend on the molar volume. The Gr\"uneisen parameter
for $\theta_p$ is the same as for $\Theta_D$. According to Edwards
and Pandorf \cite{EdwardsPandorf} for $\mathrm{^4He}$~ $\dx
\Theta_D = 7.12\times 10^4 /V_m^{2.60}~ \mathrm{K}$. For
$\theta_p$ one has
 \begin{equation}\label{eq:thetap(Vm)}
 \theta_p=
{0.91\times10^4}~{V_m^{-2.60}}~ \mathrm{K}.
\end{equation}
We plot in Fig.~\ref{fig:V20_98},  Fig.~\ref{fig:V20_84} and
Fig.~\ref{fig:V20_62} the dependencies (\ref{eq:D^-1}) and
(\ref{eq:D-Final}). The quantity $\theta_p$ is calculated using
(\ref{eq:thetap(Vm)}). The experimental points are from
Ref.~\onlinecite{AllenRichardsSchratter82}. The fit may be
improved by further variation of the parameters, but this seems
not reasonable having in mind the approximations made. It is seen
that the activation energy $\Phi$ is in good agreement with what
has been observed experimentally \cite{AllenRichardsSchratter82}
and increases with decreasing molar volume as it should. The band
width also manifests a strong molar dependence changing almost two
times from $V_m = 21~\mathrm{cm^3}$ to $V_m =
20.60~\mathrm{cm^3}$. The corresponding Gr\"uneisen parameter is
approximately 25 in good agreement with
eq.~(\ref{eq:gamma_delta}).

\begin{figure} 
\includegraphics[width=2.in]{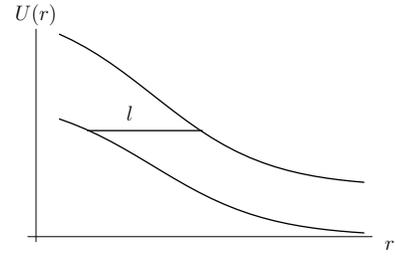}
\caption{\label{fig:BandShape} Defecton band shape in the
deformation field. The mean free path $l$ is determined by the
\emph{scattering on the defecton band boundaries}.}
\end{figure}

\begin{figure} 
\includegraphics[width=2.5 in]{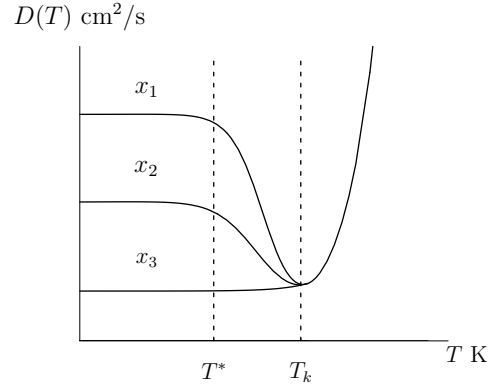}
\caption{\label{fig:SchematicD} Schematic representation of the
diffusion coefficient for $x_3<x_2<x_1$.}
\end{figure}
\begin{figure} 
\includegraphics[width=1.6in]{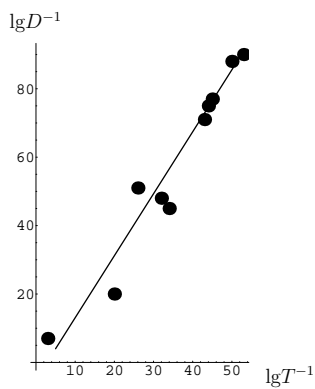}
\caption{\label{fig:T^9} The dependence  of the diffusion
coefficient on the temperature. The straight line correspond to
the law $\lg D^{-1} = 9 \lg T + 6.613$.}
\end{figure}

\begin{figure} 
 \includegraphics[width=2 in]{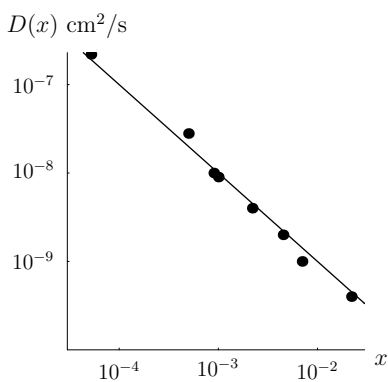}
\caption{\label{fig:D_0(x)} Concentration dependence of the
diffusion coefficient at $V_m = 20,95\ \mathrm{cm^3}$ in the
region of the low-temperature plateau after
\cite{EsselsonJETP78}.}
\end{figure}

 \begin{figure}  
 \includegraphics[width=3in]{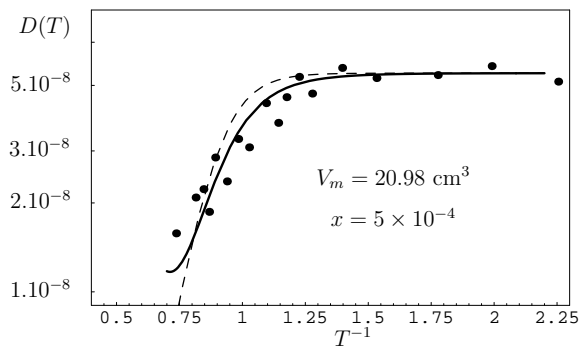}
 \caption{\label{fig:V20_98} Temperature dependence of the
diffusion coefficient at fixed concentration $x =5.10^{-4}$ and a
molar volumes $V_m = 20.98~ \mathrm{cm^3}$. Experimental points
from Ref.~\onlinecite{AllenRichardsSchratter82}. Solid line:
theoretical curve (\ref{eq:D^-1}) for  $\theta_p = 3.33~
\mathrm{K}$, $\Delta = 0.9\times10^{-4}~ \mathrm{K}$, $D_0 =
5.5\times 10^{-8}~\mathrm{cm^2/s}$,  $\omega_v = 3\times
10^9~\mathrm{s^{-1}}$ \cite{AllenRichardsSchratter82}, $\Phi =
7.1~\mathrm{K}$; dashed line corresponds to
Eq.~(\ref{eq:DKS-total}), $\Phi = 11.5~\mathrm{K}$, $\omega_v =
1.2\times 10^9~\mathrm{s^{-1}}$.}
\end{figure}

\begin{figure}  
 \includegraphics[width=3in]{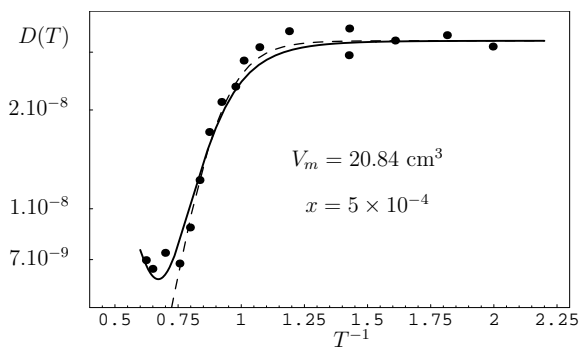}
 \caption{\label{fig:V20_84} Temperature dependence of the
diffusion coefficient at fixed concentration $x =5.10^{-4}$ and a
molar volumes $V_m = 20.84~ \mathrm{cm^3}$.  Experimental points
from Ref.~\onlinecite{AllenRichardsSchratter82}. Solid line:
theoretical curve (\ref{eq:D^-1}) for $\theta_p = 3.39~
\mathrm{K}$, $\Delta = 0.9\times 10^{-4}~ \mathrm{K}$, $\Phi =
8~\mathrm{K}$; dashed line -- Eq.~(\ref{eq:DKS-total}) with the
same parameters as in Fig.~\ref{fig:V20_98} but $\Phi =
11~\mathrm{K}$.}
\end{figure}

\begin{figure}  
 \includegraphics[width=3in]{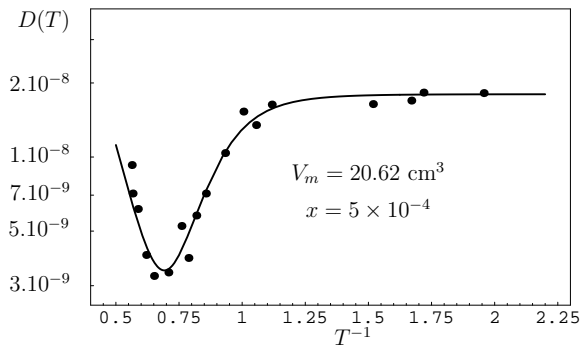}
 \caption{\label{fig:V20_62} Temperature dependence of the
diffusion coefficient at fixed concentration $x =5.10^{-4}$ and a
molar volumes $V_m = 20.62~ \mathrm{cm^3}$. Experimental points
from Ref.~\onlinecite{AllenRichardsSchratter82}. Solid line:
theoretical curve (\ref{eq:D-Final}) with $\Phi = 9~ \mathrm{K}$,
$D_0 = 1.8\times 10^{-8}$ \cite{AllenRichardsSchratter82},
$\theta_p = 3.47~ \mathrm{K}$ (\ref{eq:thetap(Vm)}), $\Delta
=0.6\times 10^{-4}~ \mathrm{K}$.}
\end{figure}
    \begin{figure}  
       \includegraphics[width=3in]{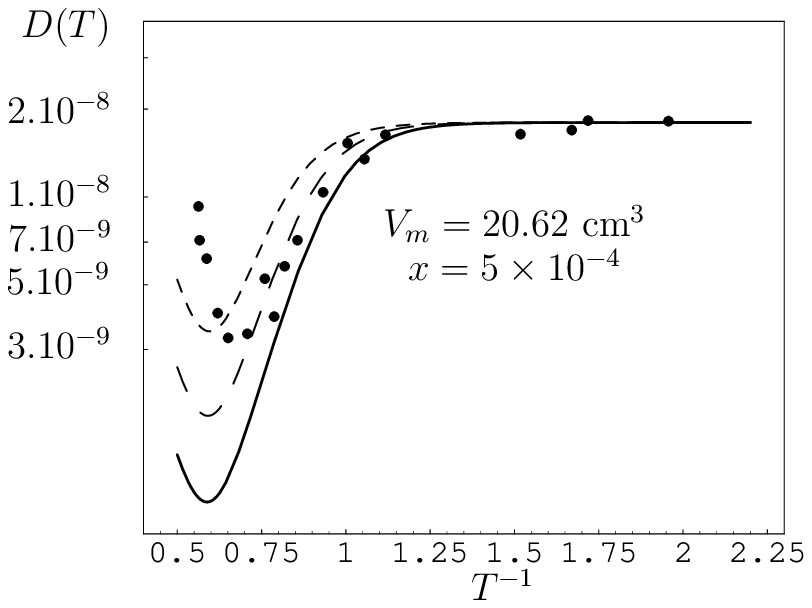}\\
    \caption{\label{fig:KSV20_62} The
diffusion coefficient (\ref{eq:DKS-total}) for fixed $\Phi =
10.5~\mathrm{K}$ and different values of $\omega$: solid line:
$\omega = 1.2 \times 10^9~\mathrm{s^{-1}}$; long dash line:
$\omega= 2.4 \times 10^9~\mathrm{s^{-1}}$; short dash line:
$\omega = 4.8\times 10^9~\mathrm{s^{-1}}$ }
    \end{figure}


\section{Phonon--controlled versus vacancy--controlled diffusion}
\label{sec:Locke}

 Although the theory of quantum diffusion has been well
confirmed by the experiment some attempts to regenerate the idea
of thermal vacancy-assisted diffusion have appeared recently
\cite{KSPRL91,KSPRB92,KSPRB93,KSReplyGri,KSReplyPush}. This
mechanism was proposed by Locke \cite{Locke78}. The diffusion
coefficient in case of vacancy-impurity scattering is of the form
(\ref{eq:D}) with $A=J_{34}^2/\omega_v$ :
\begin{equation}\label{eq:Locke}
D_{IV} = \frac{\pi}{4}\frac{k_B J_{34}^2a^2}{\hbar\omega_v
x_v\sigma_{IV}}
\end{equation}
where $\omega_v$ is the $^4$He-- vacancy exchange rate, $x_v$ is
the vacancy concentration, and $\sigma_{IV}$ is the
impurity--vacancy cross-section (in units $a^2$). Since the
vacancy concentration increases with increasing temperature as
$x_v\sim e^{-\Phi/T}$, the diffusion coefficient varies
exponentially and increases with decreasing temperature as
\begin{equation}\label{eq:D_IV(T)}
D_{IV} = D_{IV}^0 e^{\Phi/T}
\end{equation}
with
\begin{equation}\label{eq:D^0_IV}
D^0_{IV} = \frac{\pi}{4}\frac{k_B
J_{34}^2a^2}{\hbar\omega_v\sigma_{IV}} = 1.77\times 10^{-14} \
\mathrm{cm^2/s}
\end{equation}
where the following values of parameters were used in
Ref.~\onlinecite{Locke78}:
  \begin{eqnarray}\label{eq:LockeValues}
    \Phi &=& 12\ \mathrm{K},~ a = 3.67\ \AA, ~ \omega_v = 0.02\
    \mathrm{K} ,\nonumber\\
     J_{34} &=& 8\times 10^{-7}\ \mathrm{K}, \quad \sigma_{IV} =
    0.25
\end{eqnarray}
The fit was found not satisfactory (see
Ref.~\cite{Locke78,AllenRichardsSchratter82}).

Kisvarsanyi and Sullivan (KS) \cite{KSPRB93} used the same formula
with higher values of $J_{34}$
\begin{equation}\label{KSnumbers}
    J_{34}/2\pi = 2.3\times 10^5~ \mathrm{Hz}, ~ \mathrm{i.e.}~
     J_{34} =
1.1\times 10^{-5} \mathrm{K}
\end{equation}
and slightly different values of the other parameters, namely:
\begin{equation}\label{eq:KSparameters}
 \sigma_{IV} = 1.40,~ \omega_v = 1.2\times 10^9~\mathrm{s^{-1}} =
0.93\times 10^{-2}~\mathrm{K}
\end{equation}
With $a = 3.27 \times 10^{-8}~\mathrm{cm}$ this yields:
\begin{equation}\label{eq:D^KS_IV}
    D^{KS}_{IV} = 1.03\times 10^{-12} e^{\Phi/T} ~\mathrm{cm^2/s}
\end{equation}
Hence, one has for the diffusion coefficient
\begin{equation}\label{eq:Dks^-1}
    (D^{KS})^{-1} = D^{-1}_0 + (D^{KS}_{IV})^{-1}
\end{equation}
instead of (\ref{eq:D^-1}). The diffusion coefficient for the
whole temperature region (\ref{eq:D-Final}) is then replaced by
\begin{equation}\label{eq:DKS-total}
    D^{KS}_{tot} = \left(D^{-1}_0 + (D^{KS}_{IV})^{-1}\right)^{-1}
    + D_v
\end{equation}
with $D_v$ from (\ref{eq:Dclassic}).

    In their works Kisvarsanyi and Sullivan \cite{KSPRB93} (see
also Refs.~\onlinecite{KSReplyGri,KSReplyPush}) argue that
eq.~(\ref{eq:DKS-total}) describes well the experimental data of
Allen, Richards and Schratter \cite{AllenRichardsSchratter82} and
that "the overall fit for only one adjustable parameter is very
good". The value of this parameter for $V_m = 20.62~\mathrm{cm^3}$
and $x = 5 \times 10^{-4}$ was found to be $\Phi =
10.5~\mathrm{K}$. We check this in Fig.~\ref{fig:KSV20_62} (solid
line). For comparison we show the function (\ref{eq:DKS-total})
for different values of $\omega$. Unfortunately, we do not see any
satisfactory agreement. Farther on, it is confusing that in KS
theory the vacancy activation energy $\Phi$ decreases with
decreasing molar volume ($\Phi = 11.5~\mathrm{K}$ at $V_m=
21~\mathrm{cm^3}$, and $\Phi = 10.5~\mathrm{K}$ at $V_m=
20.62~\mathrm{cm^3}$). Some other comments on their theory can be
found in Refs.~\onlinecite{GriComment,PushComment}. However, there
are two points we have to discuss here. One is concerning the
impurity--vacancy cross-section $\sigma_{IV} = 1.40~\mathrm{a^2}$.
There are no details about its calculation. The only available
information is that "the standard central force approach as given
by Goldstein \cite{Goldstein}" and Lennard-Jones 6--12 potential
have been used. Following the reference link one sees that one of
the authors cites a private communication to herself
\cite{KSPRL91}. This important point, therefore, needs
elucidation.

   The second point concerns neglecting of the
impuriton-phonon scattering leading to $ D_T= G^{-1} T^{-9}$ (cf.
Eqs. (\ref{eq:T^9}), (\ref{eq:D^-1short}), (\ref{eq:B&G})). KS
have compared the reciprocal values of the diffusion coefficients
given by expressions (\ref{eq:T^9}) and (\ref{eq:D_IV(T)}) and
have come to the conclusion that the impuriton-phonon scattering
is weaker by more than two orders. As a result, they have
neglected $ D_T$. However, as it was shown in \cite{PushComment},
they have made a plain arithmetical mistake when evaluating the
factor $G$. Indeed, they have written the factor G in the form
\begin{equation} \label{eq:GKS}
G^{-1}_{KS} = \frac{5z}{32 \pi^9}(6\pi^2)^{-2/3} \frac{a^2
J_{34}^2}{
 \omega_D} \Theta_D^9 = 4.14\times 10^{-6}\frac{a^2 J_{34}^2}{
\omega_D} \Theta_D^9
\end{equation}
and have used the following set of parameters
 $$
 \Theta_D = 30~\mathrm{K},\quad \omega_D = k_B \Theta_D/\hbar =3.9 \times
10^{12}~\mathrm{s^{-1}},
 $$
 $$ J_{34} = 2\pi\times  2.3 \times 10^5 =
1.44\times 10^6~\mathrm{s^{-1}}
 $$
Substituting into (\ref{eq:GKS}) yields
 \begin{equation}\label{eq:GKSvalue}
G^{-1}_{KS} = 0.45 \times 10^{-7}~\mathrm{cm^2s^{-1}K^9},\quad
G_{KS}= 2.2\times 10^7
\end{equation}
in perfect agreement with the values used in our works and
confirmed by the experiment. KS have calculated $G^{-1}_{KS}=
6.0\times 10^{-5}$, i.e. more than $10^3$ times larger. As a
result, they have neglected the larger term, not the smaller.

In their Reply \cite{KSReplyPush} to the Comment
\cite{PushComment}, KS have made an attempt to justify themselves
by an inappropriate variation of the set of parameters. They
changed $\Theta_D$ from 30~K to 33~K (this Debye temperature
corresponds to $V_m = 19.18~\mathrm{cm^3}$ far away from the
region considered; in fact $\Theta_D = 26~\mathrm{K}$ for $V_m =
21~\mathrm{cm^3}$ \cite{EdwardsPandorf}), and $J_{34}$ from
$1.5\times 10^6~\mathrm{s^{-1}}$ to $15.7 \times
10^6~\mathrm{s^{-1}} = 1.2\times 10^{-4}~\mathrm{K}$ (a value more
than 10 times larger than the experimental one). But even with
these unrealistic values they were not able to reach $6.0\times
10^{-5}~\mathrm{K}$.


\section{Conclusion}

  We presented in this work the main ideas and results on quantum
diffusion of defects in solid helium. We showed that the strong
temperature dependence, predicted in \cite{AL69,JETP70} can be
explained very well by the phonon-defecton scattering. The
interpolation formula (\ref{eq:D-Final}) is in the excellent
agreement with the experimental data available. The theory allows
evaluation of some important quasiparticle characteristic by
direct comparison with the experiment. The exchange integral and
defecton band width were estimated. The fundamental role of the
narrowness of energy band width in the defecton-defecton
scattering and the resulting giant cros-sections predicted in
\cite{JETP70,JETPLett74} were confirmed. The dependence of the
kinetic characteristics on the molar volume was analyzed as well.
A critical analysis of some other theoretical models and
calculations was made. In particular, the inconsistency of the
vacancy-assisted diffusion \cite{KSPRB93} at low temperatures was
shown.

\begin{acknowledgments}
Acknowledgments are due to the partial support of this work from
the National Science Council, Contract F-911.
\end{acknowledgments}

\end{document}